# A High-Efficiency Reconfigurable Bidirectional Array Antenna Based on Transmit-Reflect Switchable Metasurface

Fan Qin, *Member, IEEE*, Jinyang Bi, *Student Member, IEEE*, Jiao Ma, *Student Member, IEEE*, Chao Gu, *Member, IEEE*, Hailin Zhang, *Member, IEEE*, Wenchi Cheng, *Senior Member, IEEE*, and Steven Gao, *Fellow, IEEE*

*Abstract*—This paper proposes a reconfigurable bidirectional array antenna with high-efficiency radiations and flexible beam-switching capability by employing a novel transmit-reflect switchable metasurface (TRSM). To realize the electromagnetic (EM) wave transmitted or reflected manipulation, a dedicated transmit-reflect switch layer (TRSL) with periodically soldered PIN diodes is introduced between two transmitted metasurfaces. By switching ON/OFF the embedded diodes, the TRSL performs as a mesh-type ground layer or polarization-grid layer, exhibiting a reflect or transmit property to the incident wave respectively. Further, utilizing the above TRSM configuration in conjunction with a microstrip feed antenna, bidirectional radiations are obtained at the same frequency and polarization. To further reduce the number of PIN diodes and control complexity, an enhanced TRSM using a single diode to control two unit cells is also investigated, resulting in half PIN diodes reduction. Since the bidirectional beam-switching is achieved by only controlling PIN diodes integrated in the ground plane instead of directly acting on the radiation element, which reduces insertion loss and avoids phase quantization errors, the proposed antenna can maintain a high aperture efficiency. To verify this concept, a prototype was designed, fabricated, and measured, demonstrating a successful realization of backward and forward patterns with peak gains of 22.3 and 22.1 dBi, and aperture efficiencies of 47.2% and 43.8%. The 3-dB gain bandwidths of reflected and transmitted modes are 13.7% and 12.3%. This antenna has the advantages of high gain, high aperture efficiency, simple configuration, cost-effectiveness, and flexible and digital beam control.

## I. INTRODUCTION

RECENTLY, with the emergence of various modern wireless communication technologies [1]-[5], antennas with bidirectional radiations allow for extensive coverage across upper and lower hemispheres and show potential applications for many complex communication scenarios, such as: cellular mobile communication, coal mine or tunnel relay communication, radio frequency identification (RFID) system, and wireless sensor network (WSN) [6]-[9]. Compared to directional antennas, bidirectional antennas with enhanced coverage are able to improve the response speed and antenna utilization efficiency, satisfying the demands of high-quality communication.

The conventional method to realize forward and backward radiations is to orient dual antenna array elements in opposite directions. Based on such a back-to-back arrangement, bidirectional beams can be easily generated by radiators such as microstrip patch antennas [10]-[11], dipoles [12]-[15], slot antennas [16], and leaky antennas [17]. To enhance the bidirectional gain, two groups of cascaded arrays can be combined to implement the high-gain dual-beam radiations [18]-[21]. However, due to the high insertion loss and high system complexity of the constrained feeding network, the conventional high-gain bidirectional antennas exhibit a bulky size, low efficiency, and lack of flexibility in design, which limits their broader applicability. To avoid these shortcomings, combining properties of reflectarray (RA) and transmitarray (TA) in a shared aperture, the transmit-reflect-array (TRA) offers an effective solution [22]-[24]. As a result of spatial-fed mechanism, this choice provides the benefits of low loss, flexible design, lightweight, and cost-effectiveness.

Based on whether the bidirectional radiations are generated simultaneously, TRAs can be divided into two categories, including time-division [25]-[30] and simultaneous [31]-[37] operating modes. In time-division mode, the TRA antenna behaves like a TA or RA separately, and forward or backward beam can be observed on demand independently. In [25]-[28], utilizing frequency selective surfaces (FSSs), the transmission and reflection beam conversion can be achieved at different frequency bands. By switching the orthogonal polarization states of feed source, TRAs evoke function alteration between TA and RA in response to various feed polarizations [29]-[30]. However, the usage of FSSs or multi-polarized feedings encounters a challenge for the reflected wave to maintain a consistent operating band or polarization with the transmitted wave. Besides, the beam-switching methods mentioned above, which rely on feed replacement, introduce extra complexity and would be impractical for real-world implementation.

On the other hand, bidirectional beams can be implemented at the same time in simultaneous TRA mode. By integrating

This work was supported in part by the National Key Research and Development Program under Grant 2023YFE3011502.

Fan Qin, Jinyang Bi, Jiao Ma, Hailin Zhang and Wenchi Cheng are with the School of Telecommunications Engineering, Xidian University, Xi'an 710071, China (e-mail: fqin@xidian.edu.cn; jybi@stu.xidian.edu.cn; jiaoma@stu.xidian.edu.cn; hlzhang@xidian.edu.cn; wccheng@xidian.edu.cn).

Chao Gu is with the Centre for Wireless Innovation, School of Electronics, Electrical Engineering and Computer Science, Queen's University Belfast, BT3 9DT Belfast, U.K. (e-mail: chao.gu@qub.ac.uk).

Steven Gao is with the Department of Electronic Engineering, Chinese University of Hong Kong, Hong Kong (e-mail: scgao@ee.cuhk.edu.hk).

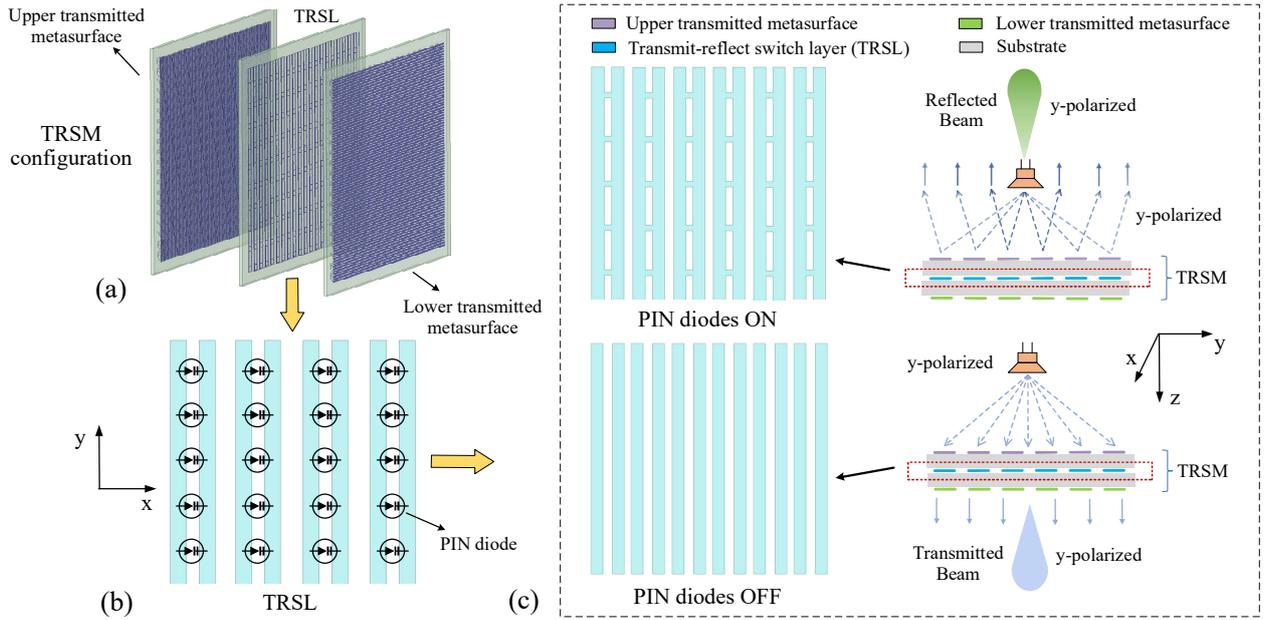

Fig. 1. (a) Configuration of the proposed reconfigurable bidirectional TRA antenna. (b) Schematic diagram of the proposed TRSL with PIN diodes integration. (c) Working concept and schematic diagram of bidirectional radiations based on the TRSM.

the transmitted and reflected unit cells into one surface to form a shared-aperture sparse array, TRAs proposed in [31]-[33] can acquire bidirectional properties simultaneously. Otherwise, when feed sources provide various polarized components simultaneously, bidirectional radiations can also be observed by TRAs [34]-[37] at the same time. However, this category has difficulty redirecting all the power in one direction at a time, leading to a relatively insufficient gain and aperture efficiency compared to the time-division approach.

So far, reconfigurable TRAs with electronic beam control have not been explored extensively. With the benefits of high response speed, flexible control, and compact integration, PIN diodes are commonly used as radio frequency (RF) switches and embedded in TRAs to realize this functionality [38]-[41]. In [40], a reconfigurable TRA with 1-bit phase quantization in both TA and RA operating modes is presented, contributing to a simultaneous or independent bidirectional beam generation. Nevertheless, the large number of RF switches exhibit a high Ohmic loss and structural complexity. Moreover, by utilizing a reconfigurable multi-polarized feed source, our previous study achieved flexible bidirectional beam switching with an insufficient aperture efficiency of 13.3% [41]. However, in all existing reconfigurable TRAs, the switches acting on the radiating aperture (involving the feed source and phase-tuning layer of the metasurface) inevitably introduce extra insertion loss or phase quantization errors, reduce radiation efficiency, and struggle to satisfy both efficient transmission-reflection mode conversion and precise phase-tuning at the same time. Hence, realizing a reconfigurable TRA with high efficiency, simple structure, especially balancing the electric switching of radiation modes and accurate phase-tuning still remains as a challenge.

This paper proposes a novel approach to achieving the bidirectional beam switching by introducing a transmit-reflect switch layer (TRSL) dedicated to PIN diode integration, without affecting the precise implementation of phase tuning and polarization rotation. Based on the TRSL, the transmit-reflect switchable metasurface (TRSM) function can alter between TA and RA with different diode states at the same operating band and polarization. To verify this concept, a TRA prototype was fabricated and measured to demonstrate good transmission and reflection properties with the benefits of high gain, high aperture efficiency, low structural complexity, and convenient electric beam control. It is worth noting that the introduced dedicated switch layer for embedding PIN diodes avoids the switches directly acting on the radiation aperture, significantly reduces the insertion loss and avoids phase quantization errors, thereby ensuring high radiation efficiency. Furthermore, an enhanced TRSM unit is proposed, which halves the number of required PIN diodes, leading to lower system complexity and improved overall performance.

The rest of the paper is organized as follows. First, the overall configuration and operating principle of the proposed antenna are elaborated. Then, a detailed discussion of the basic unit and low-complexity unit, along with the transmit-reflect switchable metasurface (TRSM) is elucidated. In Section IV, experimental results are compared, with a discussion of the comparison with existing studies. Finally, conclusions are drawn in Section V.

## II. OVERALL FRAMEWORK AND OPERATING PRINCIPLE OF THE BIDIRECTIONAL ANTENNA

The system configuration of the proposed reconfigurable bidirectional TRA antenna is illustrated in Fig. 1(a), which incorporates a novel transmit-reflect switchable metasurface (TRSM) and a wideband microstrip parasitic feeding antenna. The TRSM comprises three parts, arranged sequentially from the +z direction. Conversely, in the reflection mode, when PIN diodes are activated, a reflected pencil beam is produced in the opposite direction. In this case, the incident wave propagates sequentially through the upper metasurface, middle TRSL, and upper metasurface. Thus, by altering the operating states of the TRSL utilizing PIN diodes, the forward and backward beam-switching can be flexibly and rapidly achieved.

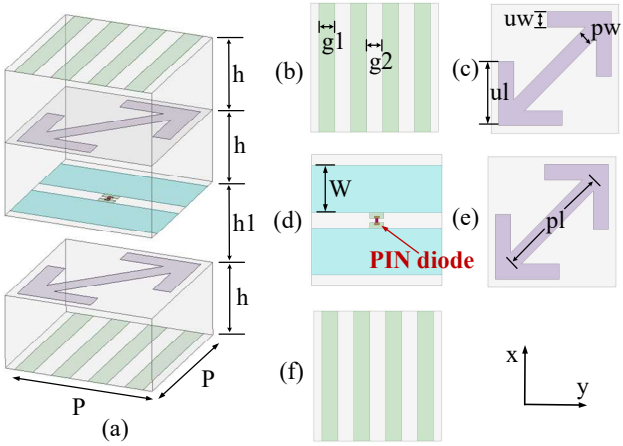

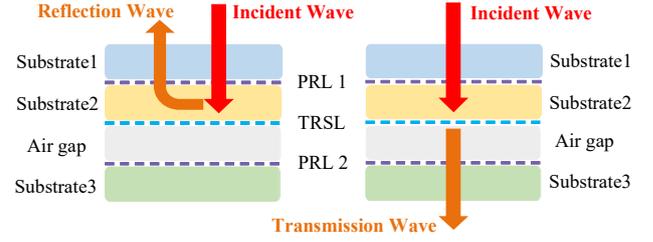

Fig. 4. Conceptual illustration of EM wave propagation when the incident wave illuminates the proposed TRSM unit cell at two modes.

## III. DESIGN OF THE PROPOSED BIDIRECTIONAL TRANSMIT-REFLECT-ARRAY ANTENNA

### A. Basic TRSM Unit Cell Design

It is worth mentioning that the transmit-reflect switch layer operates independently, allowing for phase tuning of the upper and lower metasurfaces without interference. Hence, these metasurfaces are substitutable and can be replaced by various transmitted metasurface structures. In view of our previous work, the double-arrow-shaped unit cell is chosen due to its wide bandwidth, extensive phase tuning range, and stable performance.

The exploded view of the proposed basic TRSM element is illustrated in Fig. 2(a), which consists of five metallic layers, three dielectric substrates, and an air gap. Five metal layers include two polarization-grid layers (PGLs) at the uppermost and lowermost surfaces, with two polarization-rotating layers (PRLs) positioned above and below the central transmit-reflect switch layer (TRSL). The geometrical parameters of the unit cell are listed in the caption of Fig. 2.

A reconfigurable metallic double strip-shaped resonator with a PIN diode embedded in the centre fulfills the role of the TRSL, integrated in the middle layer of the TRSM unit. With the PIN diode switched on, the connected structure, featuring a center-symmetrical and "H"-shaped configuration, functions as a mesh-type reflector for both x- and y-polarizations, effectively reflecting all the components of the incident wave. On the contrary, with the PIN diode switched off, the separated strips along the y-axis perform as y-axis-oriented polarization grids, resulting in an entire transmission of the x-polarized wave. To demonstrate the above operation, the electric field distributions of the TRSL unit under various operating modes are shown in Fig. 3. This figure illustrates a strong electric field distribution along the parallel strips when the PIN diode is switched on, whereas a weak electric field around the PIN diode and soldering pad. Oppositely, with the PIN diode switched off, a pronounced electric field appears near the soldering pad and its PIN diode on account of the top to bottom involving an upper transmitted metasurface, a middle transmit-reflect switch layer (TRSL), and a lower transmitted metasurface. The TRSL sandwiched between upper and lower metasurface apertures determines the function of TRSM to transmit or reflect electromagnetic (EM) waves illuminated by the feeding source. The upper and lower metasurfaces are engineered to provide complementary phase distributions to generate forward and backward pencil beams with high efficiency and high directivity.

Fig. 2. (a) The exploded view of the proposed basic TRSM unit cell. (b) upper polarization-grid layer. (c) upper polarization-rotating layer. (d) middle transmit-reflect switch layer. (e) lower polarization-rotating layer. (f) lower polarization-grid layer. The geometrical dimensions of the element are as follows: h = 4 mm, h1 = 4.5 mm, P = 9.2 mm, uw = 1.1 mm, pw = 1.1 mm, pl = 8.4 mm, W = 3.3 mm, g1 = 1.15 mm, g2 = 1.15 mm.

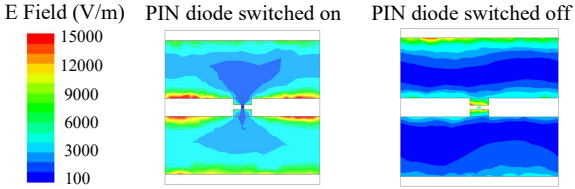

Fig. 3. The electric field distributions of the transmit-reflect switch layer. (a) reflection mode. (b) transmission mode.

As depicted in Fig. 1(b), The proposed novel structure TRSL is composed of several parallel-oriented elongated rectangular metal strips along the y-axis. PIN diodes are strategically soldered at the gaps between the metal strips of the TRSL, enabling selective functionality to either transmit or reflect the incident wave. When PIN diodes are switched on, the paralleled separated strips are connected and converted to a mesh-type structure and serve as a metal ground plane, ensuring to reflect all the components of the incident waves including both x- and y-polarizations. On the contrary, with PIN diodes switched off, the paralleled strips perform as polarization grids, allowing only the component of an incident wave with polarization along the x-axis to pass through while reflecting the y-polarized waves.

Fig. 1(c) illustrates the operating concept of the proposed TRA antenna. In transmission mode, with PIN diodes turned off, the incident wave emitted by the feed antenna propagates sequentially through the upper metasurface, middle TRSL, and lower metasurface, generating a transmitted pencil beam along

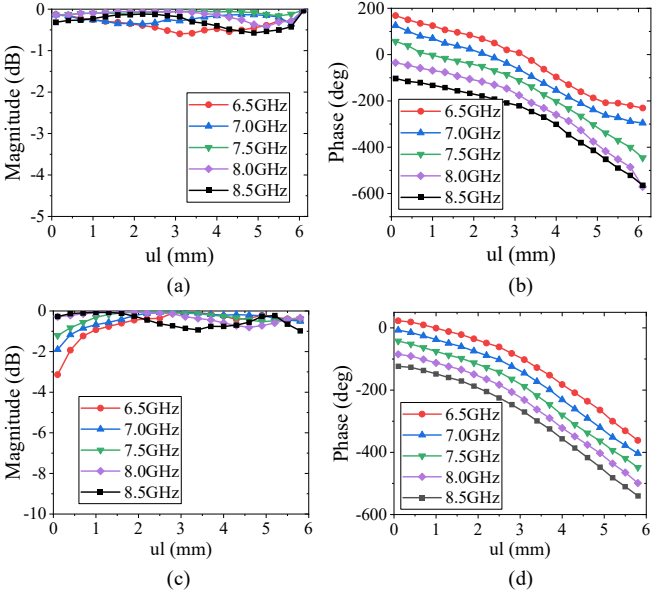

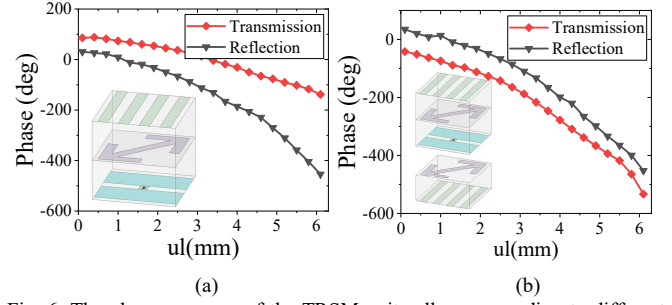

Fig. 5. Reflection and transmission magnitude and phase versus the parameter of "*ul*" at different frequencies. (a) amplitude and (b) phase of the reflection mode. (c) amplitude and (d) phase of the transmission mode.

transmitted EM wave being obstructed by the narrow spacing of the soldering pad.

Two double-arrow-shaped units with identical dimensions and phase-tuning characteristics are employed on the upper and lower sides of the TRSL, serving as the polarization-rotating layer of the reflectarray and transmitarray. By adjusting the length of parameter "*ul*", the PRL unit exhibits distinct phase-tuning properties for incident wave. Notably, since the reflected wave passes through the upper PRL twice, the variations of "*ul*" for both upper and lower PRL units are synchronized to maintain consistency, thereby ensuring an identical phase distribution between two modes to generate high-gain and high-directional beams in the +/-z directions, and effectively simplifying the overall design complexity.

Furthermore, this PRL unit is rotated with an inclined angle of 45° with respect to the x-axis, exhibiting the ability to rotate the polarization of the transmitted wave with respect to that of the incident wave by 90°. As a result, owing to the joint action of the upper and lower PRLs and polarization rotation for two times, the polarization of emergent wave remains consistent with that of the incident wave. A pair of parallel-oriented polarization grids along the x-axis, cover the outer surface of the dielectric substrate, serving as polarization-grid layers and allowing only the y-polarized waves to pass through.

The conceptual illustration of EM wave propagation, when the incident wave illuminates the proposed TRSM unit cell at transmission and reflection two modes, is shown in Fig. 4. When a y-polarized wave is incident from top to bottom of the TRSM, it initially passes through the upper PGL and is subsequently converted into an x-polarized wave by the upper PRL. This process leads to two possible outcomes:

1) *Reflection Mode:* When the PIN diode is turned on, the x-polarized wave is reflected by the mesh-type TRSL, and in turn converted back to a y-polarized wave by the upper PRL, and ultimately passes through the upper PGL. In this case, a reflected pencil beam along the -z direction can be observed.

Fig. 6. The phase response of the TRSM unit cell corresponding to different polarization-rotating layers. (a) single polarization-rotating layer media. (b) double polarization-rotating layer media.

2) *Transmission Mode:* Conversely, when the PIN diode is turned off, the x-polarized wave passes through the TRSL and is then converted back to a y-polarized wave by the lower PRL. This wave then proceeds through the lower PGL, resulting in a generation of transmitted pencil beam along the +z direction.

F4B is selected and adopted as the dielectric substrate with a relative permittivity of 2.65, loss tangent of 0.001, and 4 mm thickness. The five metal layers are typically etched or inlaid onto the surfaces of four stacked F4B substrates. However, to reserve space for PIN diode integration, this configuration can be achieved using only three dielectric substrates, with one of the substrates replaced by an air gap. Further, to ensure that the amplitude and phase between transmission and reflection modes stay consistent, the electrical size of the air gap is approximately equal to that of the thickness of a single dielectric substrate, so that the emergent waves propagate the same distance in reflection and transmission two states. As a result, this air gap is set to a thickness of h1 = 4.5mm.

To efficiently control phase and convert polarization across the desired frequency band, the parameters of the TRSM unit were optimized using a full-wave simulation software HFSS. Master-slave boundary conditions and Floquet ports were employed within the simulation process. The period of the unit is set as P = 9.2 mm, corresponding to 0.23 $\lambda$, where $\lambda$ is the free space wavelength at the center frequency of 7.5 GHz. To enhance the simulation accuracy, selecting an appropriate PIN diode and constructing an accurate equivalent circuit model are crucial. The MACOM MADP-000907-14020 PIN diode is chosen as the suitable component. For simulation of the ON state, this diode can be approximated as a resistance of 7.8 $\Omega$ and an inductance of 30 pH in series. In the OFF state, it exhibits a capacitance of 0.025 pF and an inductance of 30 pH in series.

To compensate for the phase delay resulting from the incident spherical wave, it is imperative to employ unit cells featuring a 360° phase-tuning range and high transmission and reflection efficiencies. Fig. 5 shows the relationship between the simulated reflection and transmission coefficients of the TRSM unit with variations of the parameter "*ul*" at different frequencies from 6.5 GHz to 8.5 GHz. As the parameter "*ul*" varies from 0.1 mm to 5.9 mm, a phase variation greater than 360° is attained. According to Fig. 5, the reflection coefficient of the unit cell is less than -0.1 dB at the center frequency of 7.5 GHz, and less than -0.6 dB within the entire frequency bands from 6.5 GHz to 8.5 GHz. The transmission coefficient is less than -0.9 dB during the whole frequency band. The phase shift curves for two radiations are almost parallel for frequencies from 6.5 to 8.5 GHz, showing wideband property and efficient phase compensation.

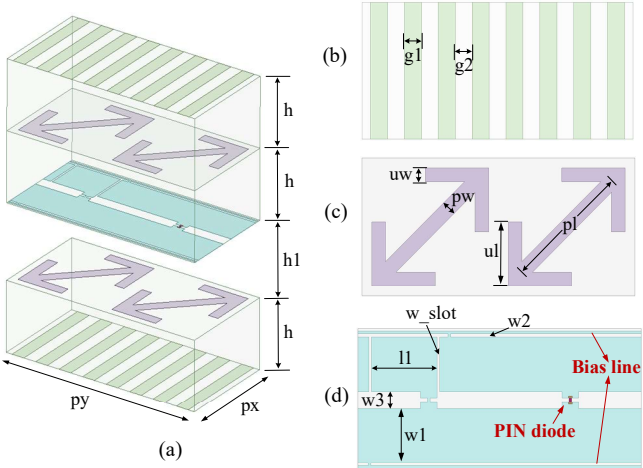

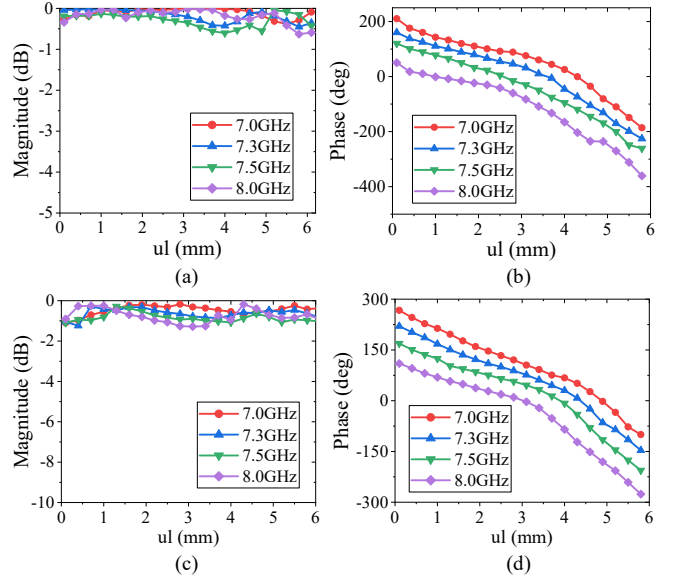

Fig. 7. (a) The exploded view of the proposed low-complexity TRSM unit cell. (b) the polarization-grid layer. (c) the polarization-rotating layer. (d) the transmit-reflect switch layer. The geometrical dimensions of the element are as follows: h = 4 mm, h1 = 4.5 mm, px = 11 mm, py = 22 mm, uw = 1.1 mm, pw = 1.2 mm, pl = 10.2 mm, w1 = 4.2 mm, w2 = 0.2 mm, w3 = 1.3 mm, l1 = 5.1 mm, w_slot = 0.2 mm, g1 = 1.375 mm, g2 = 1.375 mm.

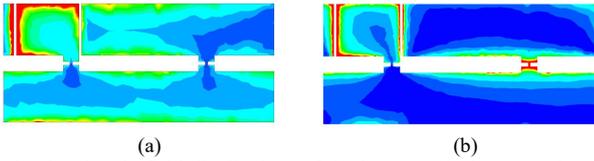

Fig. 8. The electric field distributions of the low-complexity transmit-reflect switch layer. (a) reflection mode. (b) transmission mode.

On account of the limitation imposed by the Lorentz force, the polarization-rotating layer struggles to surpass constraints of its maximum phase-tuning capability without rotating or mirroring the unit. As shown in Fig. 6(a), by varying the dimension of "*ul*", the TRSM unit with a single PRL is able to provide a 200° phase coverage. To ensure an identical phase distribution between reflection and transmission modes and achieve a phase shift range exceeding 360° simultaneously, a multi-layer TRSM unit integrating two PRLs with the same phase-tuning characteristics into a single element is employed. As shown in Fig. 6(b), this design expands the phase coverage from 200° to 400° for both transmission and reflection states, and the phase-varying tendencies are consistent between the two modes. Meanwhile, the dual-PRL TRSM unit enhances the consistency between two radiation modes, ensuring that both operating modes exhibit the same phase distribution and axial beam pointing. Since the added PRL layer is inserted behind the metal ground in the reflection mode, it minimizes the effect on the reflection performance.

Fig. 9. The amplitude and phase response of the enhanced TRSM unit cell at different frequencies corresponding to reflection and transmission properties. (a) amplitude and (b) phase of the reflection mode. (c) amplitude and (d) phase of the transmission mode.

### B. Low-complexity TRSM Unit Cell Design

Implementing the bias circuit for a pair of upper and lower polarization-rotating layer units with a single PIN diode leads to an excessive number of PIN diodes, complex control circuits, and increased costs. To address these issues, a novel and enhanced TRSM unit has been designed and optimized, allowing the control of two pairs of PRL units through a single PIN diode. This innovation reduces the number of PIN diodes used by half and effectively lowers the systematic complexity, while ensuring efficient amplitude-phase control and flexible bidirectional beam-switching as well.

The exploded view of the proposed low-complexity diode-saving TRSM element is illustrated in Fig. 7(a). The metallic double strip-shaped resonator with a single PIN diode soldered still serves as the transmit-reflect switch layer. The new unit cell design incorporates two pairs of double-arrow-shaped PRL units, comprising four units on both the upper and lower sides of the TRSL, respectively. The geometrical parameters of the unit cell are listed in the caption of Fig. 7.

The enhanced TRSL is composed of two parallel narrow metal strips with uniform spacing along the y-axis, connected by a very thin (0.2mm width) metallic wire in the left half of the unit. One PIN diode is soldered in the gap across two strips on the right half of the unit, along the x-axis. Additionally, the left portion of the upper strip is split into a short section and isolated to implement short circuit protection. To isolate the RF signal from direct current (DC), two high-impedance lines are utilized as the DC bias lines and deliver DC power to control the PIN switches. To verify this novel configuration, the electric field distributions of the low-complexity TRSL unit under various operating modes are depicted in Fig. 8, which similar to that of the basic TRSL unit cell. With the PIN diode switched on, there is a significant electric field distribution in the direction along the parallel metal grids, while perpendicular to the grids at the PIN diode connection, there is almost no electric field distributed. On the contrary, the electric field distribution is reversed when the PIN diode is switched off.

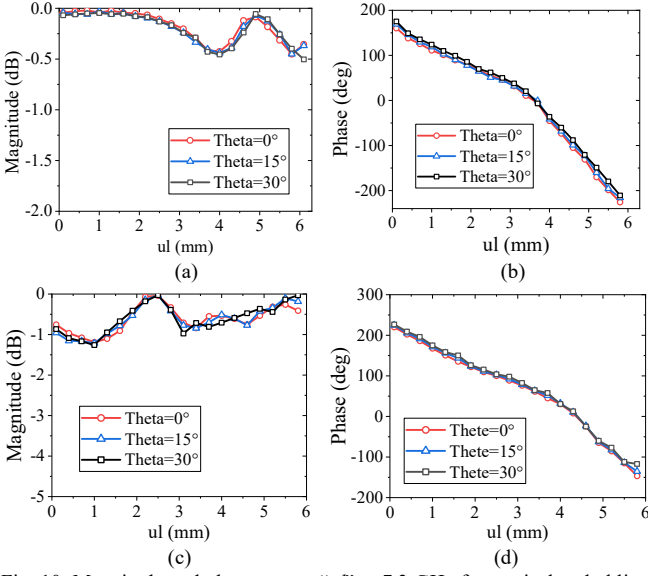

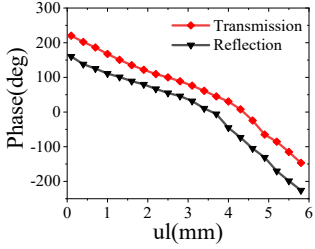

Fig. 10. Magnitude and phase versus "*ul*" at 7.3 GHz for vertical and oblique incidence with theta from 0° to 30°. (a) amplitude and (b) phase of the reflection mode. (c) amplitude and (d) phase of the transmission mode.

Fig. 11. Reflection and transmission phase curves of the enhanced TRSM unit under normal incidence at 7.3 GHz.

The reflection and transmission coefficients of the enhanced low-complexity TRSM unit for amplitude and phase under normal incidence are shown in Fig. 9 with various parameters "*ul*" from 7.0 GHz to 8.0 GHz. According to the graph, the reflection coefficient of the unit is less than -0.6 dB, and the transmission coefficient is less than -1.2 dB. The varying tendencies of the phase curves are uniform and both phase-changing ranges are greater than 360°, demonstrating the effectiveness of the improved diode-saving TRSM unit design. In Fig. 10, the amplitude and phase curves for vertical and oblique incidences at 7.3 GHz are depicted, revealing that the oblique incidence exhibits only minor amplitude and phase deviations and the performance of the unit remains stable. Analogously, a comparison of the phase between reflection and transmission modes under normal incidence at 7.3 GHz is also presented in Fig. 11, illustrating that the phase-varying tendencies are consistent between the two modes.

Although the insertion loss of the unit is slightly increased, it is worth mentioning that such a design enables utilizing a single diode to control the transmission and reflection states of four PRL units, thereby effectively reducing the number of PIN diodes used by half and the system complexity. This reduction in component count directly lowers the overall cost and simplifies the assembly process, improving the system efficiency. Moreover, it also enhances ease of integration, making it more suitable for compact and high-performance applications where minimizing complexity is crucial. In short, such a design further improves the performance of TRSM and expands a broader application prospect.

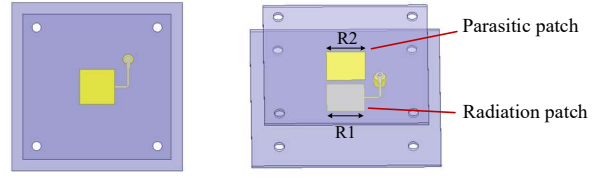

Fig. 12. The configuration of the microstrip parasitic feeding antenna.

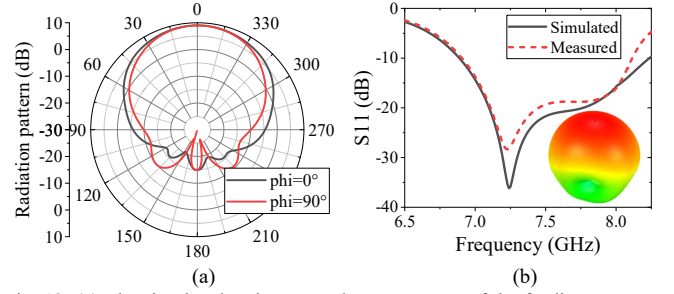

Fig. 13. (a) The simulated and measured S-parameters of the feeding antenna. (b) the simulated radiation patterns of the feeding antenna.

### C. Feeding Antenna Design

To satisfy the requirements of high-gain, wideband, and compact physical size, the feeding antenna must combine the advantages of broadband, high gain, low profile, and simple structure at the same time. A wideband microstrip parasitic feeding antenna is selected and fulfills this role.

Fig. 12 depicts the configuration of the designed microstrip parasitic antenna, which consists of two dielectric substrates of F4B with a thickness of 0.8 mm and relative permittivity of 2.65, a metal ground, a radiation patch with R1 = 11.1 mm, and a parasitic patch with R2 = 13.4 mm. The inclusion of a parasitic patch positioned above the radiation patch antenna introduces an extra resonance, enhancing its performance. Through carefully optimizing the dimensions of the parasitic patch, the bandwidth of the whole antenna can be significantly extended. Fig. 13(b) presents the simulated and measured S11 parameters of the feeding antenna, which demonstrates the bandwidth covering the frequency range from 6.90 GHz to 8.25 GHz, resulting in a relative bandwidth of 18.5%. Fig. 13(a) indicates the simulated radiation patterns of the feeding antenna. The simulated peak gain of the parasitic feeding antenna is 9.0 dBi at 7.3 GHz.

### D. Spatial Phase Compensation

The spatial phase difference arises from varying distances between each element and the phase center of the feed source. In order to achieve bidirectional high-gain radiation beams, the spatial phase difference of each element needs to be compensated in the transmission and reflection bands, and the spatial phase compensations are calculated separately as:

$$\phi(x_i, y_i) = \frac{2\pi}{\lambda_0}\sqrt{(x_i - x_f)^2 + (y_i - y_f)^2 + z_f^2} + \phi_0 \quad (1)$$

where $(x_i, y_i, 0)$ is the coordinates of the $i$th element. $\lambda_0$ is the central operating wavelength in free space. $(x_f, y_f, z_f)$ is the phase center coordinate of the feed source. $\phi_0$ is the reference phase, whose values range from 0° to 360°.

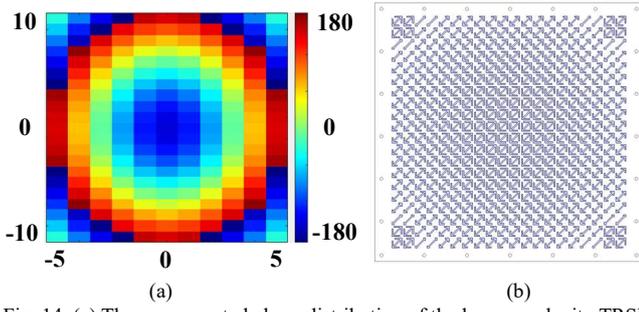

Fig. 14. (a) The compensated phase distribution of the low-complexity TRSM. (b) the polarization-rotating layer of the TRSM

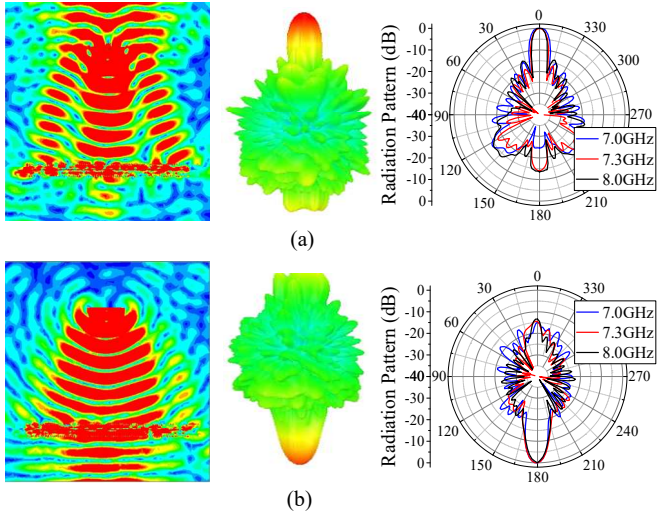

Fig. 15. Simulated E-field distributions and radiation patterns of the proposed TRA antenna. (a) reflection mode (b) transmission mode.

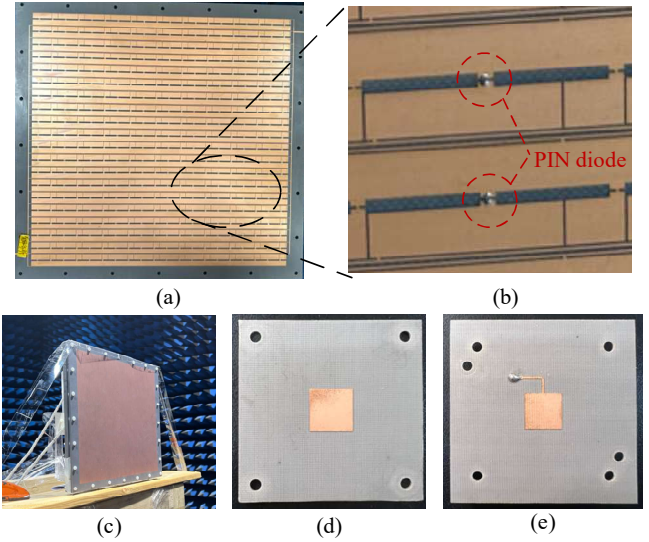

Fig. 16. The photograph of the TRSM-based TRA antenna. (a) TRSL (b) partial enlarged TRSL (c) complete assembly of the TRA antenna (d) parasitic patch of the microstrip parasitic antenna (e) radiation patch of the microstrip parasitic antenna.

The relationship between the aperture dimension, focal length, and edge taper can be defined by equation (2), where D represents the lateral size of the antenna and F represents the focal length. To implement an optimum aperture efficiency, the focal length-to-diameter ratio F/D is determined by using the $\alpha$−10dB (-10 dB taper of the feed antenna) to obtain the optimal aperture efficiency. This approach ensures the best possible aperture efficiency by appropriately balancing the feed illumination and minimizing spillover losses.

$$f = \frac{D}{2\tan(\alpha_{-10dB})} \quad (2)$$

Based on the above design theory, the focal length of F is optimized to 150.4 mm. The F/D ratio is set as 0.72. In order to ensure that the transmitted and reflected waves form highly directional pencil beams along the normal directions, the phase compensation distribution of the transmission mode is set equal to that of the reflection. According to (1) and (2), the transmission and reflection phase distribution is calculated as shown in Fig. 14(a). After calculating each element phase in two radiation states, the phase-shifting parameter "$ul$" of each element can be derived using the data presented in Fig. 9. The upper and lower polarization-rotating layers with identical configuration are depicted in Fig. 14(b).

To demonstrate the operation of the bidirectional beams, the electric E-field distributions and radiation patterns for forward and backward radiations are shown in Fig. 15. The TRSM effectively converts quasi-spherical waves into plane waves, and the far-field patterns confirm that the antenna generates bidirectional beams with forward radiation in the transmission state and backward radiation in the reflection state.

## IV. SIMULATION AND MEASUREMENT RESULTS

To validate the antenna design strategy, a reconfigurable bidirectional TRA antenna with 20 × 10 low-complexity TRSM elements is fabricated and measured in a microwave anechoic chamber, as depicted in Fig. 16. Figs. 16(a) and (b) show the metal patterns of the inner transmit-reflect switch layer with PIN diodes embedded. Fig. 16(c) illustrates the complete assembly of the antenna. The microstrip parasitic feed source is depicted in Figs. 16(d) and (e). Nylon screws are utilized to fasten the multi-layer metasurface structure, and acrylic frames with greater hardness are selected to support the feed source and minimize their coupling with the antenna. The effective aperture size of the antenna is 220 mm × 220 mm × 163 mm (5.35 λ × 5.35 λ ×3.97 λ).

To implement the reconfigurable operation, 200 PIN diodes are embedded to control the TRSM states through bias lines. Table I lists the radiation characteristics of the two operating states. Correspondingly, the far-field radiation patterns were measured in an anechoic chamber using a standard horn. The measured normalized radiation patterns for the bidirectional antenna combined with simulated ones at the frequency points of 7.0, 7.3, 7.7, and 8.0 GHz are plotted in Figs. 17 and 18. It is observed that the transmitted and reflected pencil beams are well generated in the upper and lower hemispheres and remain a wideband frequency range from 7.0 to 8.0 GHz.

TABLE I
WORKING STATES AND RADIATION CHARACTERISTICS OF THE PROPOSED BIDIRECTIONAL TRA ANTENNA

| Feed polarization | Operating states | PIN diodes working states | Main beam direction | Polarization of generated beam |
|---|---|---|---|---|
| y-polarization | Reflection | on | Upper hemisphere | y-polarization |
| | Transmission | off | Bottom hemisphere | |

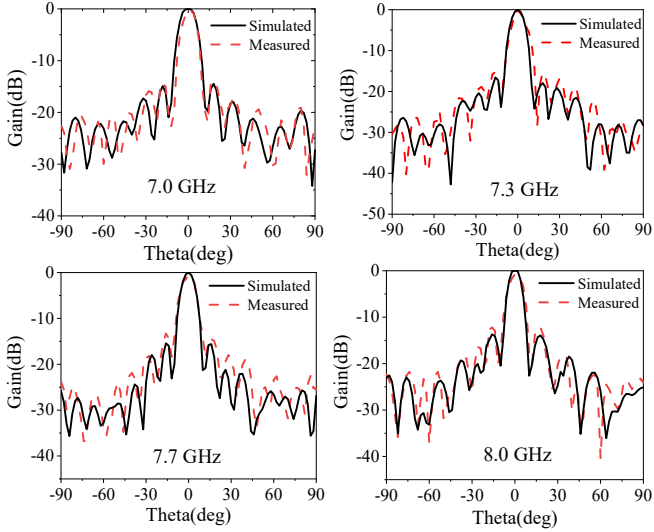

Fig. 17. Radiation patterns of the TRA antenna in the reflection mode.

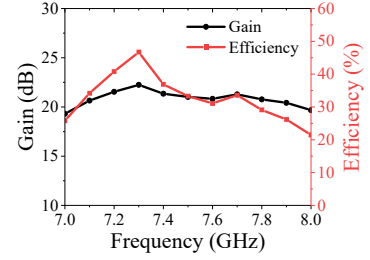

Fig. 19. Measured gains and aperture efficiencies of the bidirectional beams in the reflection mode

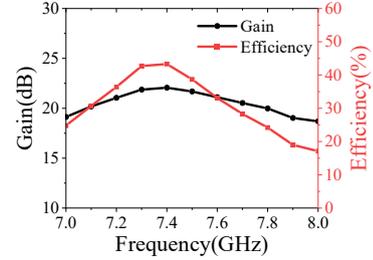

Fig. 20. Measured gains and aperture efficiencies of the bidirectional beams in the transmission mode.

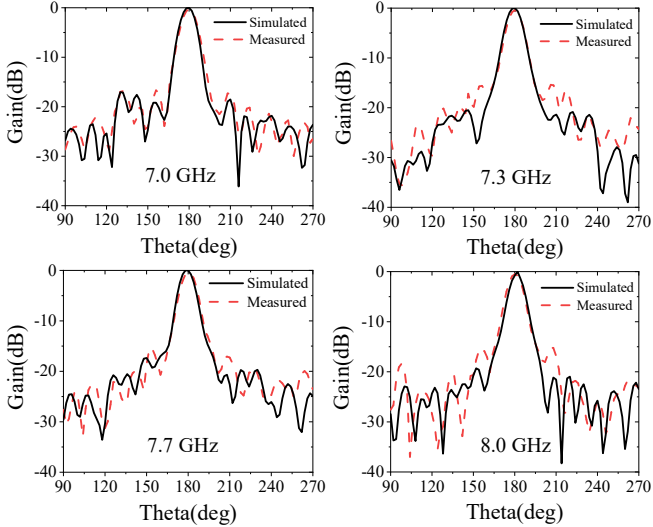

Fig. 18. Radiation patterns of the TRA antenna in the transmission mode.

Meanwhile, the measured gains and aperture efficiencies of the antenna with different operating states and frequencies are also investigated in Figs. 19 and 20. For the reflected beam, the simulated maximum gain is 22.9 dBi at 7.3 GHz. The measured peak gain is 22.3 dBi at 7.3 GHz, corresponding to a peak aperture efficiency of 47.2%. For the transmitted beam, the simulated peak gain is 22.7 dBi at 7.3 GHz. The measured maximum gain is 22.1 dBi at 7.4 GHz, corresponding to a peak aperture efficiency of 43.8%. The observed discrepancies between the measured and simulated results are mainly attributed to substrate losses and assembly inaccuracies. The measured 3-dB gain bandwidth for the reflected beam is 13.7%, spanning from 7.0 to 8.0 GHz, while the transmitted beam achieves a bandwidth of 12.3%. The gain performance is stable, less than 3.6 dB, across the entire operating bandwidth. The measured results agree well with the theoretical results in terms of gain, bandwidth, and efficiency.

The introduction of an additional transmit-reflect switch layer dedicated to achieving the bidirectional beam switching, instead of relying on PIN diodes loaded on the radiating apertures such as the polarization-rotating layer of metasurface, allows for independent control of functions involving phase tuning and bidirectional beam-switching. The utilization of a separated switchable layer significantly reduces quantization errors and facilitates bidirectional mode conversion without adversely affecting phase, enabling more precise and efficient beam control. Consequently, the measured aperture efficiency is maintained at a relatively high level.

To highlight the novelty and advantages of the proposed array, a comparison with the published bidirectional antennas is summarized in Table II. Unlike prior research, our design

TABLE II
COMPARISON WITH PROPOSED RELATED TRA ANTENNAS

| Ref | Radiation Type | 3-dBGain bandwidth (%) | Beam-switching | Aperture size | Peak Gain (dBi) | Peak AE (%) |
|---|---|---|---|---|---|---|
| [22] | T+R | 15 / 14 (1dB) | NO | 14λ × 14λ | 25 / 25.5 | 14 / 15 |
| [38] | T+R | 32 / 32 | NO | 7.1λ × 7.1λ | 15.4 / 17.2 | 5.4 / 8.2 |
| [25] | T/R | / | Feed replacement | 9λ × 9λ | 20.8 / 21.9 | / |
| [28] | T/R | 7.4 / 8.1 (1dB) | Feed replacement | 7.5λ × 7.5λ | 23.1 / 21.3 | 29.1 / 32.1 |
| [32] | T+R | 6.7 / 9.3 (1dB) | NO | 12.5λ × 12.5λ | 21.4 / 24.4 | 7 / 14 |
| [40] | T/R, T+R | 19.5 / 19.9 | PIN diodes | 7.6λ × 7.6λ | 21.4 / 21.0, 17.1 / 16.2 | 19.2 / 17.3, 7.1 / 5.7 |
| **This work** | **T/R** | **12.3 / 13.7** | **PIN diodes** | **5.4λ × 5.4λ** | **22.1 / 22.3** | **43.8 / 47.2** |

*T refers to transmitarray; R refers to reflectarray; AE refers to aperture efficiency;

facilitates flexible beam-switching capability between the forward and backward radiation modes, maintaining the same polarization and frequency between the two directional beams. By introducing an additional TRSL layer with PIN diodes embedded to realize the beam-switching, the proposed novel TRSM shows the enhanced reconfigurable performance of transmitting and reflecting the incident wave with low loss, resulting in properties of high aperture efficiency and high gain of the array. Meanwhile, this layer ensures that the bidirectional beam-switching and phase-tuning remain independent, preventing any interference between these two functionalities. Its adaptable design allows it to seamlessly integrate with various types of transmit-reflect switchable metasurfaces, making it highly versatile for combination with a wide range of other metasurface structures. The benefits of this design involving a high gain, high aperture efficiency, simple configuration, and flexible digital control ensure a wide application prospect in wireless communication systems.

To the authors' knowledge, this approach, utilizing a TRSM metasurface with an extra electric switch layer, is applied in bidirectional antenna design to achieve high efficiency and convenient beam-switching for the first time.

## V. CONCLUSION

In this paper, we present a novel structure of the transmit-reflect switchable metasurface with capability of bidirectional beam manipulation. By introducing an additional metal layer with PIN diodes embedded specialized for beam switching, the TRSM implements flexible and independent reflection and transmission mode conversion with high efficiency. Based on this function selection configuration, a prototype has been parametrically optimized and measured to reach forward and backward pencil beams along the +/-z direction with the same polarization and frequency. For the reflected beam, the measured peak gain is 22.3 dBi, corresponding to a peak aperture efficiency of 47.2%. For the transmitted beam, the measured maximum gain is 22.1 dBi, corresponding to a peak aperture efficiency of 43.8%. The proposed antenna features high gain, high aperture efficiency, low structural complexity, and straightforward beam control. Meanwhile, the specialized switch layer design is well-suited for integration with a diverse array of transmitted metasurfaces, enabling its application across a broad spectrum of use cases.